\documentclass[a4paper]{jpconf}
\usepackage[utf8]{inputenc}
\usepackage{amsmath}
\usepackage{amsfonts}
\usepackage{amssymb}
\usepackage{color}
\usepackage{graphicx}
\usepackage{hyperref}
\usepackage{lastpage}
\usepackage{fancyhdr}
\newcommand{\nn}{\nonumber}
\begin{document}
\title{Alternative Description of Magnetic Monopoles in Quantum Mechanics}

\author{Samuel Kov\'a\v{c}ik}

\address{Dublin Institute for Advanced Studies, 10 Burlington Road, Dublin 4, Ireland}

\ead{skovacik@stp.dias.ie}

\author{Peter Prešnajder}

\address{Faculty of Mathematics, Physics and Informatics, Comenius University Bratislava, Mlynsk\'a dolina, Bratislava, 842 48, Slovakia}

\ead{presnajder@fmph.uniba.sk}

\begin{abstract}
We present an alternative description of magnetic monopoles by lifting quantum mechanics from 3-dimensional space into a one with 2 complex dimensions. Magnetic monopoles are realized by generalization of the considered states. Usual algebraic relations and magnetic fields describing monopoles are reproduced, with the Dirac quantisation condition satisfied naturally.
\end{abstract}

Ref. number:  DIAS-STP-18-04

\section{Introduction}
Magnetic monopoles are a vital part of many theoretical models, despite never being actually observed. They have a long stretching history – in classical electromagnetism they appeared as a generalization of Maxwell equations (by adding magnetic charge sources). The situation is more intricate in quantum mechanics, where electromagnetism is described by electromagnet potentials, which were defined to make the theory free of magnetic monopoles by default (as $\mbox{div rot } \textbf{A} \sim \rho_M = 0$). 

However, there is a workaround – one can use potentials singular on a (half-)line, so called Dirac strings, \cite{dirac}. This idea allowed Dirac to study magnetic monopoles in the context of quantum mechanics (QM) and to derive the famous Dirac quantisation condition, which states that the product of electric and magnetic charge, denoted $\mu$, has to be quantized. Therefore, if at least one magnetic monopole exists in the Universe the electric charge has to be quantised – as is indeed observed. 

This description was later refurbished by Yang and Wu \cite{YangWu, Yang}, who instead of using one singular potential used two non-singular potentials defined in different, but overlapping, regions. This connected the theory of magnetic monopoles not only to the mathematical theory of sections, but also, in restrospective, to topology. Two potentials define two solutions which are related by a phase factor transformation. On the overlapping region, one can define a non-contractible path and by moving along the entire loop, the phase has to change by an integer multiple of $2\pi$. The winding number integer corresponds to the quantised $\mu$. Important studies of magnetic monopoles in QM, whose results we are comparing with in this paper, were done by Zwanziger, see \cite{zwanziger}. 

Nowadays, magnetic monopoles often appear as topological solution in various field theories, where they appear as long as one can specify topologically nontrivial boundary conditions (for example the hedge-hog configuration in \cite{pol, hooft} or in super Yang-Mills theory in \cite{Seiberg:1994rs}). They also appear in the string theory, where they are even less similar to the electric ones, as their dimensionality usually differs \footnote{The condition is $p+p' = D - 4$, see \cite{Gauntlett:1992nn}, where the electrical object is $p-$dimensional, the magnetic one is $p'-$dimensional and $D$ is the dimension of spacetime. In the currently observed universe is $D=4$ and therefore $p=p'=0$.}.

In this paper we will show how can be the QM monopoles described in a new way. The main idea is this – we will reformulate QM in $\textbf{C}^2$, a space with two complex dimensions, instead of $\textbf{R}^3$. We will show that the usual QM can be obtained by restricting on a specific Hilbert subspace. Then, we will show that by lowering these restrictions we can describe magnetic monopoles of an arbitrary charge $\mu$.

That the monopoles can be described in $\textbf{C}^2$ using a Hopf fibration is known, but mostly from the context of electromagnetism, see \cite{Ryder, Minami}. Our approach is closely related to QM and can be used as a starting point for some other modifications, for example for formulating QM on a non-commutative space, see \cite{Jabbari, LRL, mm, mm2}.

\section{Quantum mechanics revisited}

Quantum-mechanical theory consist of two building blocks: a Hilbert space of states and a set of operators on it. The most notorious example might be the space of square-integrable complex functions $\Psi(x)$ of three spatial coordinates $x_i, i =1,2,3$, which is equipped with a norm 
\begin{equation} \label{norm}
||\Psi || ^2 = \int \Psi^* (x) \Psi(x) d^3x .
\end{equation}
Observables are defined as operators on this space, for example the coordinate and the momentum operators are
\begin{eqnarray}
\hat{x}_i \Psi(x) &=& x_i \Psi(x), \\ \nn
\hat{p}_i \Psi(x) &=& - i \partial_i \Psi(x),
\end{eqnarray}
where we have set $\hbar=1$. As long as normalizable under \eqref{norm}, any function $\Psi(x)$ can be considered a QM state. 

This entire picture can be recast in a different way, instead of starting from a space with three real coordinates $x_i$, we can take two complex coordinates $z_\alpha, \alpha =1,2$. The Hilbert state consists of square-integrable complex functions $\Phi(z,z^*)$ and is equipped with a norm
\begin{equation} \label{norm1}
||\Phi^2 || = \int \frac{2 r}{\pi} \Phi^* (z,z^*) \Phi(z,z^*) dz d\bar{z},
\end{equation}
where $r\,=\,  \bar{z}_\alpha z_\alpha$ (summation over repeated indexes is assumed in this paper). The choice of this weight will become clear later.
The space $\textbf{C}^2$ is naturally equipped with a symplectic structure
\begin{equation} \label{Pois}
\{ z_\alpha ,z^*_\beta \}\ =\ - i\,\delta_{\alpha\beta}\,,\ \  \{ z_\alpha ,z_\beta \}\ =\ \{ z^*_\alpha ,z^*_\beta \}\ =\ 0.
\end{equation}
Using this we can define numerous operators on the Hilbert space, for example the Laplace operator
\begin{equation} \label{lapl}
\Delta \Phi(z,z^*) =\frac{1}{r}\,\{z^*_\alpha, \{z_\alpha,\Phi(z,z^*)\}\}.
\end{equation}
So far this looks like completely different QM, but there is a way to connect the theory in $\textbf{C}^2$ to that in $\textbf{R}^3$. The important thing to recall is that even though these two spaces are different, the groups of their (rotational) symmetries are locally isomorphic (both can be realized using the algebra of Pauli matrices). We can parametrise $z_\alpha, z^*_\alpha$ using the Euler angles as
\begin{eqnarray}\label{euler}
z_1 =& \sqrt{r} \cos \left(\theta /2 \right) e^{\frac{i}{2}\left(-\phi+\gamma \right)}, \ z^*_1 &= \sqrt{r} \cos \left(\theta /2\right) e^{-\frac{i}{2}\left(-\phi+\gamma \right)}, \\ \nonumber \label{CP}
z_2 =& \sqrt{r} \sin \left(\theta /2 \right) e^{\frac{i}{2}\left(\phi+\gamma \right)}, \ z^*_2 &= \sqrt{r} \sin \left(\theta /2 \right) e^{-\frac{i}{2}\left(\phi+\gamma \right)} . 
\end{eqnarray}
The coordinates of $\textbf{C}^2$ can be mapped into the coordinates of $\textbf{R}^3$ using the Pauli matrices as
\begin{equation} \label{Hopf}
x^i = \bar{z} \sigma^i z.
\end{equation}
How does this map work? It maps a point from a 3-sphere $\bar{z} z = r$ into a point on a 2-sphere $x^2 = r^2$. All points differing only the angle $\gamma$ are mapped into a single one. This is a (complex) Hopf fibration $S^3 \stackrel{S^1} {\rightarrow} S^2$ for any $r>0$. Plugging \eqref{euler} into \eqref{Hopf} results into $x^i$ expressed in the spherical coordinates $(r,\theta, \phi)$.

Instead of considering any square-integrable function $\Phi(z,z^*)$ as a QM state, we can restrict only to functions depending on a very specific combinations of $z_\alpha, z_\alpha^*$, namely those of the form $\Phi(x)$ with $x$ defined in \eqref{Hopf}. Then, the naturally defined operators act in a familiar way
\begin{eqnarray} \label{ope}
 \Delta \Phi(x) &=& \frac{1}{r} \{z^*_\alpha, \{z_\alpha,\Phi(x)\}\} = \partial_{x_i} \partial_{x_i} \Phi(x), \\ \nn
 \hat{x}_i \Phi(x) &=& x_i \Phi(x), \\ \nn
 \hat{V}_i \Phi(x)&\equiv &\frac{1}{2}\left[ \Delta, \hat{x}_i\right] \Phi(x) \\ \nn
 &=& - \frac{i}{2r}\sigma ^i_{\alpha \beta} (z^*_\alpha \partial _{z^*_\beta} + z_\beta \partial_{z_\alpha}) \Phi(x) \\ \nn
 &=& -i \partial_{x_i} \Phi(x), \\ \nn
 \hat{L}_i \Phi(x) &=& \frac{i}{2}\{ x_i, \Phi(x) \} = \varepsilon_{ijk} \hat{x}_j \hat{V}_k \Phi(x) .
\end{eqnarray}
Also, the integration weight becomes $\int \frac{r}{2\pi} d^2z \rightarrow \int d^3x$. The third relation defines the velocity operator, which is equal to the conjugate momentum (with $m = \hbar = 1$) and the fourth one defines the angular momentum operator satisfying the usual $su(2)$ relation. All of these can be obtained using \eqref{Pois} and the chain rule for derivatives (note that $\{z_\alpha , . \} = -i \partial_{z^*_\alpha}  $ and $\{z^*_\alpha , . \} = i \partial_{z_\alpha} $). 

This way we can rewrite the standard $\textbf{R}^3$ QM in $\textbf{C}^2$ formalism. By considering only states of the form $\Phi(x)$ (restring on a certain Hilbert subspace) it is hard to see any difference. 

One of the benefits of this construction is that the symplectic structure \eqref{Pois} allows for Kontsevich quantisation of the underlying space, resulting into a theory of QM on a noncommutative (or quantum, fuzzy) space $\textbf{R}^3_\lambda$, \cite{Jabbari, LRL}. We will not follow this direction here, it can be found in \cite{mm, mm2}.

For the considered Hilbert subspace with states of the form $\Phi(x)$ is the singularity at $r=0$ only a coordinate one. However, for general states this point has to be removed as it would make the Laplace operator $\Delta$ defined ill-defined. From now on, we consider $\textbf{C}^2 \setminus \{0\}$ instead of $\textbf{C}^2$.

\section{Generalized states}
The important property of the states $\Phi(x)$ is that they contain an equal powers of complex coordinates $z$ and their conjugates $z^*$ which is ensured by the form of \eqref{Hopf}. As a result, their dependence on the angle $\gamma$ disappears.

Let us now do the following: We will lower the restriction for the form of the states, allowing them to have an unequal powers of $z$ and $z^*$. These states have no counter-part in the ordinary QM, as they do depend on $\gamma$. Nonetheless, the operators \eqref{ope} remain well-defined and we will use them to show that the generalized states actually described magnetic monopoles of an arbitrary charge allowed by the Dirac quantisation condition.

The generalized states forming $\mathcal{H}_\kappa$ are of the form
\begin{equation} \label{kStates}
\Phi_\kappa(z,z^*) = \Phi(x) \cdot \xi_\kappa, \ \xi_\kappa = \left( \frac{z_1}{z^*_1} \right)^{\frac{\kappa - \delta}{4}} \left( \frac{z_2}{z^*_2} \right)^{\frac{\kappa + \delta}{4}},
\end{equation}
or expressed in coordinates \eqref{euler} as
\begin{equation}
 \xi_\kappa = e^{i \frac{\kappa}{2} \gamma} e^{i \frac{\delta}{2} \phi}.
\end{equation}
We could consider even a sum of such terms with different $\delta$ (and different coefficients), but will stick with this simplest case. As the coordinates $\gamma, \phi$ are periodic with a period of $4\pi$, the functions $\Phi_\kappa$ remain uniquely defined as long as $\kappa, \delta \in \textbf{N}$. While $\kappa$ counts the difference in the powers of $z$ and $z^*$, $\delta$ counts the difference between the components with index $1$ and $2$. 

It is important to notice that while for $\kappa \neq 0$ we cannot express the wave-function $\Phi_k$ as a function of $x$, the probability density $\Phi_k^* \Phi_k$ always contains equal powers of $z$ a $z^*$ and can be, via \eqref{Hopf}, expressed as a function of $\textbf{R}^3$ coordinates. Therefore, the states always have proper probabilistic interpretation. 

There are many relations that reveal the monopole behavior, for example the following relation 
\begin{equation}
\varepsilon_{ijk} \hat{x}_j \hat{V}_k \Phi_\kappa=\left( \hat{L}_i + \frac{\kappa}{2} \frac{\hat{x}_i}{r} \right)\Phi_\kappa.
\end{equation}
reveals presence non-vanishing angular momentum. The same was shown for a monopole system by Zwanziger in \cite{zwanziger}. Probably the most telling object is the commutator of the velocity operators
\begin{equation}
[\hat{V}_i, \hat{V}_j]\Phi_\kappa = \left(+ \frac{\kappa}{2} \right) i \varepsilon_{ijk} \frac{\hat{x}_k}{r^3}\Phi_\kappa,
\end{equation}
which is equal to the monopole field strength (also the same as in the Zwanziger study). 

This should be sufficient, but we can proceed one step further. The velocity operator is a differential operator, acting only on $\Phi(x)$, it is proportional to the partial derivative with respect to $x^i$. However, its action on $\Phi_\kappa$ is more sophisticated, as by the Leibniz rule there is also a contribution $\Phi (x) \hat{V}_i \xi_\kappa$. Let us identify it with a gauge potential
\begin{equation} \label{comVA}
\hat{V}_j \Phi_\kappa = (- i \partial_{x^j} \Phi(x))\xi_\kappa + A_j \Phi_\kappa, A_j= - \frac{i}{2r\xi_\kappa} \sigma ^j _{\gamma \delta} z_\delta (\partial _{z_\gamma} \xi_\kappa)
\end{equation} 
Given our choice of $\xi_\kappa$, the only nontrivial component (in spherical coordinates) is
\begin{equation}
A_\phi = \frac{\delta + \kappa \cos (\theta)}{2r \sin (\theta) },
\end{equation}
which defines the following magnetic fields
\begin{equation}
B_i = (\mbox{rot } \textbf{A})_i  = -\frac{\kappa}{2} \frac{x_i}{r^3} .
\end{equation}
As we can see, the resulting Coulomb-like field depends only on $\kappa$, not $\delta$. The potential $A_\phi$ can be split into two parts, the first one $A_\phi^\kappa = \frac{\kappa}{2r}\cot(\theta)$ describes the magnetic field while the other, $A_\phi^\delta = \frac{\delta}{2 r }\csc (\theta) = \nabla_\phi \left(\frac{\delta}{2}\phi \right)$ can be gauged away. Both of them appear in the analysis of Yang, see \cite{Yang}. 

It is obvious now that $\kappa/2$ can be identified with the magnetic charge $\mu$. Dirac quantisation condition that $\mu$ has to be a half-integer is satisfied naturally as $\kappa$ counts the difference in the powers of $z$ and $z^*$, which is in our construction an integer. 

If we set $\delta=0$ we have $\xi_\kappa = \left(\frac{z_1 z_2}{z_1^* z_2*}\right)^{\frac{\kappa}{4}}$ and the resulting potential $A_\phi$ is singular for both $\theta = 0, \pi$. On the other hand choosing $\delta = \pm \kappa$ yields $\xi_\kappa = \left(\frac{z_1}{z_1^*}\right)^{\frac{\kappa}{2}}$ or $\xi_\kappa = \left(\frac{z_2}{z_2^*}\right)^{\frac{\kappa}{2}}$, which leads to a potential singular only on the north or the south pole. Therefore we conclude that $\left(\frac{z_1}{z_1^*}\right)$ and $\left(\frac{z_2}{z_2^*}\right)$ correspond to Dirac semi-strings with $\mu=\frac{\kappa}{2}$ and the general choice \eqref{comVA} is a combination of both. 

It is also possible to define additional operator $\hat{V}_4 = \frac{1}{2 r} \left( z^*_\alpha \partial_{z^*_\alpha} - z_\alpha \partial_{z_\alpha} \right)$ which acts as $\hat{V}_4 = \frac{1}{r}\partial_\gamma$ on the considered states and measures their monopole charge.

\section{Conclusion}

We have shown a slightly unusual construction of QM. The three-dimensional theory was reformulated in four-dimensional (or two-complex-dimensional) space by considering a restricted class of wave-functions of the form $\Phi(\bar{z} \sigma z)$. This construction is alluring as it offers some new possibilities. As we have shown, the extra dimension, which in the context of Hopf fibrations has a topology of a circle, allowed us to describe monopoles states in a regular way. 

Also, as this space is naturally equipped with Poisson structure, it can be (canonically) quantised to obtain noncommutative space: $ \{ z_\alpha, \bar{z}_\beta \} = - i \delta_{\alpha \beta} \rightarrow [z_\alpha , z_\beta^+] = \delta_{\alpha \beta}$ and $[x_i, x_j] = 2 i \varepsilon_{ijk} x_k$. Such spaces, whose close points cannot be distinguished below some fundamental scale, are a rather general aspect of theories of quantum gravity.

After this quantisation, one can define QM the same way as was done in this paper to obtained so-called noncommutative QM, see \cite{Jabbari, LRL,groenewold,snyder, fuzzy}. By considering a similarly generalized class of states as in this paper (containing an unequal number of bosonic creation and annihilation operators), one can realize magnetic monopoles in this theory, see \cite{mm, mm2}. 

The fact that lifting the theory one dimension up offered new possibilities shall be of no surprise – when Kaluza and Klein studied general relativity in 5 dimensions, with the fifth dimension tightly curled up, electromagnetism fitted in in a very natural way. Also not so surprisingly, there are known gravitational solutions in 5 dimensions that look like magnetic monopoles in 4 dimensions, \cite{Sorkin:1983ns, Gross:1983hb}. 

Our choice of the form of $\xi_k$ ensured that the factor which introduced monopoles was just a phase, not affecting the norm of the states $||\Phi_\kappa||^2 = || \Phi (x) \xi_k ||^2 = || \Phi (x) ||^2$. If we were more careless and chose, for example, $\xi_k = (z_1 z_2)^{\frac{\kappa}{2}}$, following the same steps we would lead to the same monopole field $B_i = -\frac{\kappa}{2} \frac{x_i}{r^3}$. The fact that such choice changes the $(r, \theta, \phi)$ dependence of the wave-functions would manifest itself as an imaginary contribution to he potential $A_i$ that can be removed easily. In fact, if we factorize $\xi_\kappa$ as $\xi'_\kappa(r,\theta,\phi) \xi''_\kappa(\gamma)$ it follows that

\begin{equation}
Im(A_i) = -  \partial_{x_i} \log (\xi'_\kappa),
\end{equation}
which, being a gradient, can be gauged away – resulting into a change of the wave-function which is not just a phase, but instead

\begin{equation}
e^{- \log( \xi_\kappa ')} \Phi(x) \xi_\kappa' \xi _\kappa'' = \Phi(x) \xi_\kappa'',
\end{equation}
taking the undesired part away, leaving only the factor $\xi_\kappa''$ that depends only on $\gamma$. 

In other words: if we introduce monopoles by adding a factor that is not just a phase, it can be gauged away using the imaginary potential it produces. Notably, complex electromagnetic potentials has been studied in the context of electro-magnetic duality in the theory with superluminal sources in \cite{Deriglazov:2016mhk}, but in our case is the imaginary part redundant.

\subsection*{Acknowledgment}
This research was partially supported by COST Action MP1405(S.K. and P.P), project VEGA 1/0985/16 (P.P.) and the Irish Research Council funding (S.K.).

\end{document}